# Comparison of Australasian tektites with Australasian microtektites and BeLaU spherules recovered from the ocean

Eugenia Hyung[1], Emma Levy[1], Loralei Cook[1], Stein B. Jacobsen[1], Abraham Loeb[2], Jayden Squire[3], Juraj Farkas[3]

[1]Department of Earth and Planetary Science, Harvard University, Cambridge, 02138, MA
[2]Department of Astronomy, Harvard University, Cambridge, 02138, MA
[3]Department of Earth Sciences, School of Physics, Chemistry and Earth Sciences, The University of Adelaide, 5000, Australia

**Abstract**

The Australasian strewn field covers more than 15% of Earth's surface, consisting of tektites and microtektites. Australasian tektites from Southeast Asia and Australia, as well as microtektites recovered from deep sea sediments and Antarctica, are established to be derived from upper continental crust sediments. An expedition to retrieve remnants of bolide CNEOS 2014 January 8 (IM1), held in the Pacific Ocean, was in proximity to the known extent of the Australasian strewn field, and yielded "BeLaU"-spherules, whose compositions did not match most well-studied solar system material. We therefore report precise and comprehensive elemental data for Australasian tektites to compare their elemental abundances to those of Australasian microtektites from deep sea sediments, and BeLaU. Our findings corroborate previous studies that Australasian tektites and microtektites closely resemble the elemental abundance patterns of the upper continental crust. Meanwhile, the elemental patterns of the BeLaU-spherules are distinct from the Australasian tektite/microtektite compositions.

1. Introduction

Tektites are melt-droplets solidified from melt or condensed from vapor due to meteoritic impacts and are characterized as glassy silicate objects that are round or oblong in shape, and dark brown or black in color. They have been established to originate from impacts onto areas with close to average upper continental crust in composition (Mizera et al. 2016 and references therein), with traces of the impactor apparent in Co, Ni, and Cr abundances, if present (e.g., Glass et al. 2004). Microtektites are tektites that are less than 1 mm in diameter, lack microlites (e.g., Glass and Simonson 2012), and are often found in marine sediments or Antarctica. Tektites and microtektites identified belonging to the Australasian strewn field have been well-studied for their chemical compositions (e.g., Glass et al. 2004, Mizera et al. 2016), where comprehensive chemical analyses prove to be insightful.

An expedition was held (Loeb et al. 2024) to retrieve remnants of a bolide named CNEOS 2014 January 8 (IM1), suggested to be of interstellar origin, due to its arrival velocity of more than 45 km/s (Siraj and Loeb 2022). Among the various objects retrieved were over 800 particles that were 0.05–1.3 mm in diameter. While the majority (~80%) of these materials consisted of cosmic spherules, classified into "S-type," "I-type," and "G-type" spherules of chondritic origins, a fraction of the materials that did not fit into the three archetypal categories were further classified as "D-type" spherules or particles. They were named after their highly differentiated compositions in comparison to chondrite, characterized by their low Mg/(Mg+Fe) contents. A small subset of the D-type particles was further subcategorized and named "BeLaU," after their unusual elemental patterns exhibited in high abundances in highly incompatible elements such as Be, La, and U compared to most known terrestrial materials, such as the continental crust. The "BeLaU" composition was proposed to yet be of unknown origin (Loeb et al. 2024).

Desch (2024a,b; 2025) suggested the BeLaU spherules are part of the Australasian tektite strewn field and are microtektites of lateritic soil. Among the various types of tektites, Australasian tektites and microtektites are a viable candidate for the BeLaU composition due to the proximity of the sample retrieval expedition site to the extent of the strewn field identified thus far for the Australasian tektites, covering parts of China, Indonesia, the Pacific Ocean, the Indian Ocean, Australia, and Antarctica (e.g., Di Vincenzo et al. 2021). The impact is dated to have happened at 0.79 Ma (e.g., Schwarz et al. 2016). The sampling site is also in proximity to the putative impact site for the Australasian tektites, widely suggested to be in Indochina (Mizera et al. 2016; Sieh et al. 2020).

We report new and precise elemental abundances of Australasian tektites for 55 elements and compare them to the elemental abundances of Australasian microtektites from deep ocean sediments (Folco et al. 2018), and the "BeLaU" compositions retrieved from the IM1 site.

## 2. Samples and analytical methods

Four Australasian tektites were chosen for analysis. The tektite samples came from the Tate Museum, University of Adelaide, where two originated from Florieton, South Australia, one from Charlotte Waters (Northern Territory of Australia), and one from Kalgoorlie (Western Australia). The samples were characterized to be glassy, homogeneous, and dark brown/black and oblong or rounded in shape. About 50–100 mg of each specimen was broken and finely crushed with a sapphire mortar and pestle and dissolved in a mixture of concentrated acids consisting of HF-$HNO_3$, followed by a dry down and dissolution in a 1:4 mixture of water and aqua regia, to be dried down once more, and finally dissolved in dilute HCl. A small aliquot was sampled from each of the dissolutions to be measured for their elemental abundances using an iCAP TQ triple quadrupole ICP-MS (ThermoFisher Scientific), using 10 ppb In-spiked 2% $HNO_3$ solutions to correct for instrumental drift.

## 3. Results and Discussion

The elemental abundances of the four tektites, and average upper continental crust (UCC) are plotted in the order of most incompatible to least compatible elements for 41 elements (elements that have well-defined positions in the compatibility series) and are normalized with respect to the primitive mantle (PM) composition of McDonough and Sun (1995) in **Figure 1a**. The same four tektite data normalized to the UCC is shown in **Figure 1b**, but in this case the elements (55 elements) are ordered according to increasing atomic number. The BeLaU composition and the average tektite composition are also plotted in this figure for direct comparison. Finally, the averaged BeLaU composition and the average composition of Australasian microtektites from deep sea sediments (Folco et al. 2018) is normalized with respect to the average of the four Australasian tektites measured in this study in **Figure 1c**, in the manner of **Figure 1b**.

The elemental abundances of the tektites are all very similar with respect to one another (**Figure 1a**), exhibiting enrichment in the highly incompatible elements, and depletion towards more compatible elements in comparison to the PM. The patterns are also similar with respect to the UCC, closely resembling the abundances and depletions in PM-normalized elemental abundances. In comparison to the UCC, the tektites are depleted in Na and P and exhibit variability in Ca among the specimens.

The UCC-normalized plot (**Figure 1b**) corroborates previous studies that tektites are generally composed of terrestrial sediments derived from the continental crust (e.g., Glass et al. 2004). Variations in Cd and Sb abundances among the tektite samples are apparent. With respect to the REEs, the tektite compositions are slightly enriched in comparison or approach near-unity to UCC. In addition to Na and P, depletions are apparent for elements such as Cu, Zn, As, Mo, Sb, Tl, Pb, and Bi. Depletions in Tl, Pb, and Bi have been observed for Australasian tektites (Taylor and McLennan 1979), interpreted to be due to volatile loss during impact. The BeLaU pattern and Australasian tektites exhibit opposite normalized enrichment patterns for Mo, with the former enriched, and the latter depleted.

The average BeLaU composition is plotted alongside the tektites in **Figure 1b**. In comparison to the tektites, UCC-normalized patterns indicate enrichment in heavy REE compared to the continental crust. While concentrations of Be and U of the BeLaU composition exhibit enrichment compared to the UCC, those of tektites are more similar to the UCC. In **Figure 1c**, a direct comparison of Australasian tektites, microtektites, and BeLaU are made through tektite-normalized elemental abundances. Despite residence in the ocean, Australasian microtektites retrieved from deep sea sediments exhibit similar elemental abundances to those of Australasian tektites, while both are distinct from BeLaU.

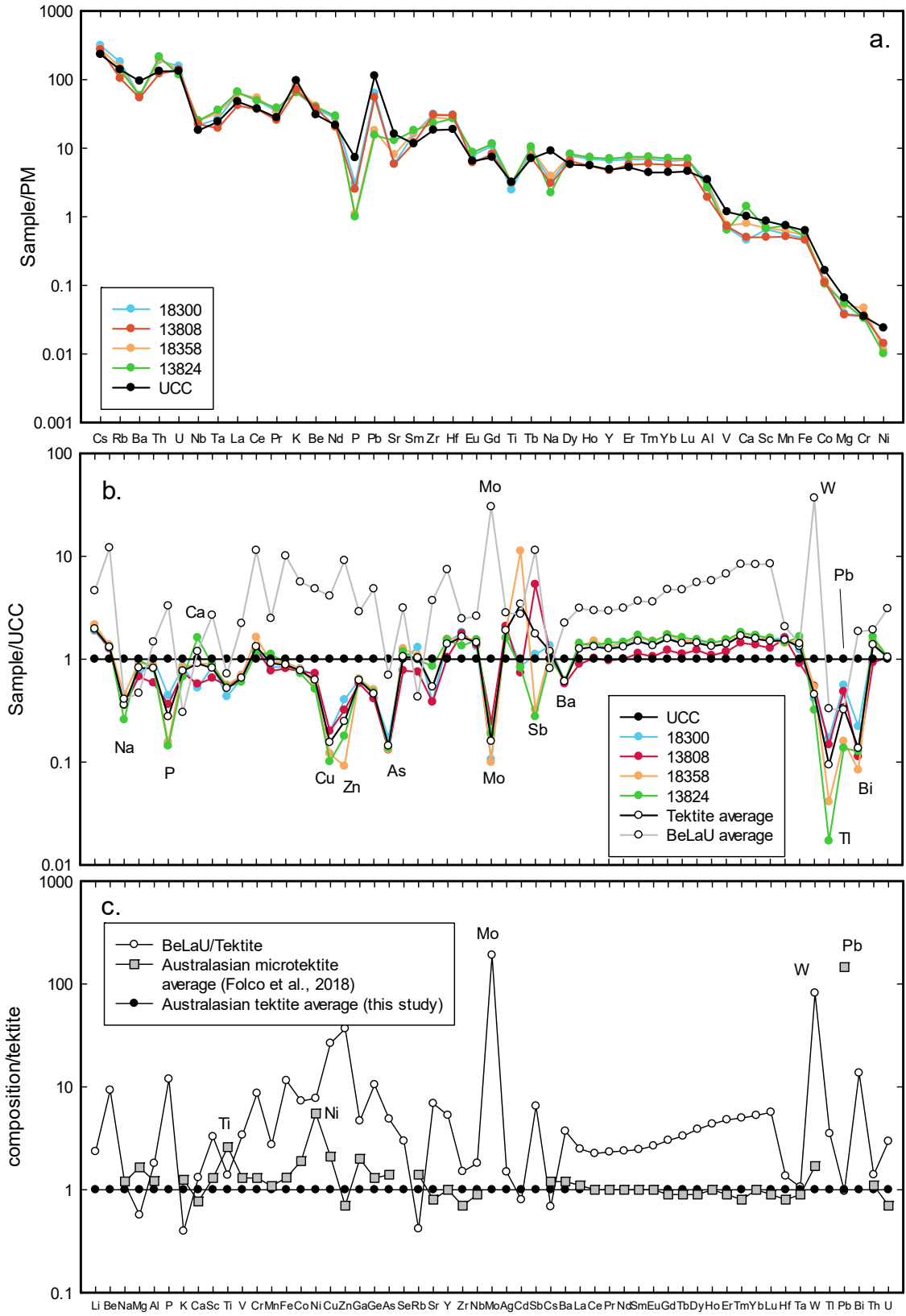

**Figure 1.** Panel (a) Primitive-mantle normalized (using McDonough and Sun 1995 values) elemental compositions of Australasian tektites compared to the UCC (Rudnick and Gao 2014), arranged in order of most incompatible to most compatible elements. Panel (b): The UCC-normalized elemental patterns of the individual tektites (red, green, blue, and orange) and the average of the four specimens (white symbols, black line). The number "1" represents the UCC (dark symbols), plotted alongside the average BeLaU abundance pattern (white symbols, grey line) normalized with respect to the UCC. Panel (c): Australasian-tektite normalized elemental compositions of BeLaU (white circles) and microtektites (grey squares; Folco et al. 2018). Here, the number "1" (black circles) represents the average of the Australasian tektites measured in this study.